\documentclass[prl,twocolumn,amsmath,amssymb,showpacs]{revtex4}

\usepackage{graphicx}
\usepackage{dcolumn}
\usepackage{bm}

\begin{document}


\title{Pair-wise decoherence in coupled spin qubit networks}

\author{ Andrea Morello$^{1}$, P. C. E. Stamp$^{1,2}$, and Igor S. Tupitsyn$^{1,2,3}$ }

\affiliation{$^1$ Department of Physics and Astronomy,
University of British Columbia, Vancouver B.C. V6T 1Z1, Canada.\\
$^2$Pacific Institute for Theoretical Physics, Vancouver B.C. V6T 1Z1,
Canada. \\
$^3$ Russian Federal Research Center ``Kurchatov Institute'',
Kurchatov Sq.1, Moscow 123182, Russia. }


\begin{abstract}
Experiments involving phase coherent dynamics of networks of spins,
such as echo experiments, will only work if decoherence can be
suppressed. We show here, by analyzing the particular example of a
crystalline network of Fe$_8$ molecules, that most decoherence
typically comes from pairwise interactions (particularly dipolar
interactions) between the spins, which cause `correlated errors'.
However at very low $T$ these are strongly suppressed. These results
have important implications for the design of quantum information
processing systems using electronic spins.

\end{abstract}


\maketitle

A worldwide effort is presently on to make nanoscale solid-state
qubits, whose purity and reproducibility can easily be controlled.
Microscopic spins, existing in molecular magnets
\cite{troiani05PRL}, quantum dots \cite{loss98PRA} and
semiconductors \cite{kane98N}, or in doped fullerenes
\cite{morton06NP}, are a leading candidate for this. In some of
these systems (notably magnetic molecules), the individual qubit
properties are controlled by chemistry instead of by
nano-engineering (the `bottom-up' approach \cite{joachim00N}), with
spin Hamiltonians and inter-molecular spin couplings known and
controlled to at least 3 significant figures. Spin also possesses
other advantages - information can be encoded in the topological
spin phase, with no need to move electrons around. Using spins for
quantum information will ultimately require (i) detecting and
manipulating single-spins, and (ii) understanding and controlling
decoherence.

\begin{figure}[b]
\includegraphics[width=8.5cm]{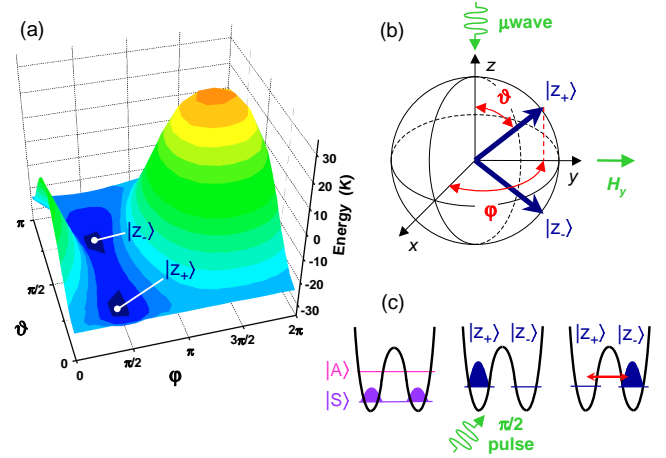}
\caption{(Color online) In a strong transverse field ${\bf H}_{\perp}$, the two potential wells of an easy axis
spin system approach each other on the Bloch sphere. (a) The spin anisotropy energy for an Fe$_8$ molecular spin
with ${\bf H}_{\perp}$ along $\hat{y}$, easy axis $\hat{z}$ and hard axis $\hat{x}$, when $\mu_{\rm o} H_y = 2.5$
T. The low-lying states $|\mathcal{Z}_{\pm}\rangle$, are approximately localized in the two potential wells. The
quantum-mechanical eigenstates are symmetric and antisymmetric superpositions of $|\mathcal{Z}_{\pm}\rangle$ (see
text), separated by the tunneling gap $2\Delta_{\rm o}$. (b) The resonance experimental set-up and the spin
states on the Bloch sphere; (c) At $T \ll \Delta_{\rm o}/k_{\rm B}$ only the lowest-energy eigenstate is
populated, $\vert {\cal S} \rangle = 2^{-1/2}(|\mathcal{Z}_{+}\rangle + |\mathcal{Z}_{-}\rangle)$. A short
$\mu$wave pulse prepares the system in the $|\mathcal{Z}_{+}\rangle$ state ($\pi/2$ rotation, corresponding to
$1/4$ of a Rabi oscillation). The spin then tunnels coherently between $|\mathcal{Z}_{+}\rangle$ and
$|\mathcal{Z}_{-}\rangle$ at a frequency $2 \Delta_{\rm o}/\hbar$. The effect of static inhomogeneities in
$\Delta_{\rm o}$ can be compensated by a $\pi$-pulse in a spin-echo sequence (not shown).} \label{l-spins}
\end{figure}

\begin{figure}[t]
\includegraphics[width=8.5cm]{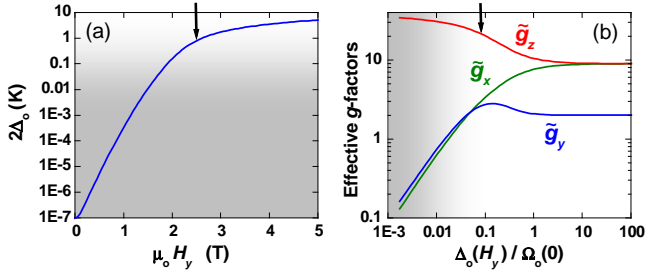}
\caption{(Color online) Fe$_8$ in a transverse field $H_y$: (a) Tunneling splitting $2 \Delta_{\rm o}$ of the
lowest doublet, as a function of $H_y$. (b) Effective $g$-factors for Fe$_8$ as a function of the ratio
$\Delta_{\rm o}(H_y)/\Omega_{\rm o}$. In both figures, the values at $\mu_{\rm o} H_y = 2.5$ T are indicated by a
black arrow. Our decoherence rate calculations are valid when $\Delta_{\rm o} \gg U_{\rm d}$, and do not apply in
the grey areas shown.}
 \label{l-geff}
\end{figure}

Single spins have been detected in a few ingenious experiments
\cite{rugar04N}, but we don't yet have a general-purpose,
single-spin detection/manipulation tool, analogous to single atom
STM/AFM. Consider, however, an array of spins, each having a
low-energy doublet of states whose splitting is easily controlled by
a magnetic field. Even without addressing individual spins, one can
still demonstrate coherent qubit operation, using external AC fields
to promote resonant transitions between levels, and pulse sequences
(e.g. spin echo) to manipulate the phase and measure decoherence
rates. This approach is well known for room-temperature bulk NMR
quantum computing \cite{gerschenfeld97S}. Here we treat the case of
electronic spins which, unlike nuclei, can be highly polarized at
low $T$. We introduce a formalism allowing the description of any
set of spin qubits obtained by truncation to low energy of a larger
system, showing how the low-$T$ decoherence rate can be dramatically
reduced, even for a network of mutually coupled qubits. To be
specific, we treat the case where the qubit is obtained by taking an
anisotropic high-spin nanomagnet \cite{gatteschi94S} with easy axis
$\hat{z}$, subject to a large transverse field ${\bf H}_{\perp}$
(Fig. \ref{l-spins}), giving a low-energy doublet of states with
easily controllable energy separation, $2\Delta_{\rm o}({\bf
H}_{\perp})$ [Fig. \ref{l-geff}(a)]. To make quantitative and
testable predictions, we then calculate the spin-echo decay rate in
a network of Fe$_8$ molecules. This is a clean, crystalline and
stoichiometric chemical compound \cite{gatteschi94S,wieghardt84AC},
where the inter-qubit and the qubit-environment interactions are
known accurately, and it should be a good `benchmark' for
quantitative test of the theory.

(i) {\it Effective Hamiltonian}: In a transverse field ${\bf
H}_{\perp}$ the effective spin Hamiltonian of the Fe$_8$ molecule,
with total spin $S=10$, is controlled by crystal and external
fields:
\begin{equation}
{\cal H}({\bf S}) = -DS^2_z + E S^2_x + K^{\perp}_4
(S^4_{+}+S^4_{-}) - \gamma {\bf S} \cdot {\bf H}_{\perp},
\label{HS10}
\end{equation}
The easy/hard axes are along $\hat{z}$ and $\hat{x}$ respectively; here $D/k_{\rm B} = 0.23$ K, $E/k_{\rm B} =
0.094$ K, $K^{\perp}_4/k_{\rm B} = -3.28 \times 10^{-5}$ K, $\gamma = g_e \mu_{\rm B} \mu_{\rm o}$ and $g_e
\approx 2$ is the isotropic $g$-factor of the spin-10 moment \cite{tupitsyn02B}. Henceforth we orient the
external field along $\hat{y}$, ie., $\mathbf{H}_{\perp} \rightarrow {\bf H}_y$; this can tune the tunneling
splitting $2 \Delta_{\rm o}$ between the two lowest eigenstates over seven orders of magnitude [Fig.
\ref{l-geff}(a)]. Higher excited states are separated from the lowest doublet by a large gap $\Omega_{\rm o}
\approx 5$ K, and for $k_{\rm B} T \ll \Omega_{\rm o}$ the giant spin can be truncated to an effective
spin-$1/2$.

Given the two lowest doublet eigenstates,
$|\mathcal{S}(\mathbf{H}_{\perp})\rangle$ and
$|\mathcal{A}(\mathbf{H}_{\perp})\rangle$, we write the
quasi-localized states in the minima of the energy potential as
$|\mathcal{Z}_{\pm}\rangle = 2^{-1/2}(| \mathcal{S} \rangle \pm |
\mathcal{A} \rangle)$ [Fig. \ref{l-spins}(a)]. Defining the states
$|\mathcal{X}_{\pm}\rangle = 2^{-1/2}(| \mathcal{S} \rangle \pm i |
\mathcal{A} \rangle)$, we see that $|\mathcal{X}_{\pm}\rangle$ and
$|\mathcal{Z}_{\pm}\rangle$ have the maximum (positive or negative)
spin expectation values along $\hat{x}$ or $\hat{z}$ within the
2-dimensional subspace of the qubit, where $|\mathcal{S}\rangle$ and
$|\mathcal{A}\rangle$ are the basis states. All of these Fe$_8$ spin
states can be calculated by numerical diagonalization of
(\ref{HS10}). We then define define qubit spin-$1/2$ operators
$\hat{s}_x, \hat{s}_y, \hat{s}_z$, such that $\hat{s}_z
|\mathcal{S}\rangle  = 1/2 |\mathcal{S}\rangle$, etc.

To describe the magnetic moment of the truncated spin qubit we introduce an effective $g$-tensor $\tilde{\bf g}$,
also operating in the qubit subspace, and defined so that the qubit magnetic moment \cite{moment} $m^s_{\mu}
(\mathbf{H}) = \mu_{\rm B} \sum_{\nu} \tilde{g}_{\mu\nu}(\mathbf{H}) s_{\nu}$. In the geometry studied here, with
${\bf H}$ along $\hat{y}$, $\tilde{\bf g}$ is diagonal, with components:
\begin{subequations}
\begin{eqnarray}
\tilde{g}_x = g_{e}(\langle \mathcal{X}_+|S_x| \mathcal{X}_+ \rangle
- \langle \mathcal{X}_-|S_x| \mathcal{X}_- \rangle); \\
\tilde{g}_y = g_{e}(\langle \mathcal{S} |S_y| \mathcal{S} \rangle
- \langle \mathcal{A} |S_y| \mathcal{A} \rangle); \\
\tilde{g}_z = g_{e}(\langle \mathcal{Z}_+|S_z| \mathcal{Z}_+ \rangle
- \langle \mathcal{Z}_-|S_z| \mathcal{Z}_- \rangle). \label{g(E)}
\end{eqnarray}
\end{subequations}
Numerical evaluation of $\tilde{\bf g}$ shows it to be highly anisotropic and field-dependent [Fig.
\ref{l-geff}(b)].

Consider now two spin-10 Fe$_8$ molecules. The standard dipolar interaction $\frac{1}{2} \sum_{i \neq j}
\sum_{\mu,\nu} U_{\mu \nu}^{ij} S_{\mu}^i S_{\nu}^j$ has strength $U_{\rm d} = \mu_{\rm o} g^2_e \mu^2_{\rm B}
S^2/4 \pi {\cal V}_{\rm c} = 0.127$ K between nearest neighbors (here ${\cal V}_{\rm c}$ is the volume of the
unit cell). However we are interested in the effective interaction between the {\it qubits} - this acquires a
field-dependent and highly anisotropic tensor form $\tilde {\bf V}^{ij}({\bf H}_{\perp}) = \tilde{\bf g}^i {\bf
U}^{ij} \tilde{\bf g}^j$ when written in the truncated qubit basis. There may also be exchange interactions
\cite{wernsdorfer02N} between the molecules - these however have never been observed in Fe$_8$. Thus our low-$T$
Hamiltonian for the dipole-interacting molecules becomes:
\begin{equation}
H_{\mathrm{eff}} = - \sum_i 2 \Delta_{\rm o}^i({\bf H}_{\perp}) s_z^i +
\frac{1}{2} \sum_{i \neq j} \sum_{\mu, \nu = x,y,z} \tilde{V}_{\mu
\nu}^{ij}({\bf H}_{\perp}) s_{\mu}^i  s_{\nu}^j, \label{H_eff}
\end{equation}
where the spins form a triclinic lattice \cite{wieghardt84AC}, and
we choose the qubit $\hat{z}$ axis of quantization to be along the
field (ie., along the original $\hat{y}$ axis). The full Hamiltonian
also includes local spin-phonon and hyperfine couplings, whose
detailed form we will not need here.

(ii) {\it Decoherence}: Most general discussions of decoherence in qubit systems concentrate on `1-qubit'
decoherence, ie., that coming from the interactions of individual qubits with the environment
\cite{leggett87RMP,weiss99,prokofev00RPP}. In an insulating magnetic system like Fe$_8$ both nuclear spins and
phonons will contribute to 1-qubit decoherence \cite{stamp04PRB}. However this is not the only possible
decoherence source. In a multi-qubit system one can have `correlated errors', from pairwise qubit interactions. A
few analyses of this have been done \cite{alicki98}; depending on what model is chosen, these indicate that when
qubits couple to the same bath, correlated decoherence is very serious, and may prevent error correcting codes
from operating.

In the set-up imagined here, decoherence will show up in measurements of the dephasing time, $T_2$. We define the
dimensionless decoherence rate as $\gamma_{\phi} = \hbar/T_2 \Delta_{\rm o}$ (the `coherence Q-factor' is then
$Q_{\phi} \sim \pi/\gamma_{\phi}$). We start by considering the role of dipolar interactions - these are
important because they exist in all spin systems, and cause correlated errors via pair-wise spin interactions,
exciting internal modes of the spin system (ie., no external environment). There are two ways to look at their
contribution to $\gamma_{\phi}$. First, as a dephasing from dipole-mediated pair-flip processes, which in a
resonance or echo experiment gives a homogenous absorption linewidth $\langle \delta \omega_{\phi}\rangle =
T_2^{-1}$. Second, as a scattering of the uniform spin precession mode off thermal magnons. In what follows we
will assume that $U_{\rm d} / 2 \Delta_{\rm o} \ll 1$, ie., the dipolar interaction $\ll$ the tunneling splitting
- only then will dipolar decoherence be small enough to make an experiment worthwhile (in Fe$_8$ at, eg.,
$\mu_{\rm o} H_y = 2.5$ T we have $U_{\rm d} / 2 \Delta_{\rm o} \approx 0.16$). Since any reliable measurement of
the decoherence rate will involve time-resolved spin-echo experiments, we neglect static inhomogeneous broadening
in the calculations.

If $U_{\rm d} / 2 \Delta_{\rm o} \ll 1$ and $k_{\rm B}T \gtrsim
\Delta_{\rm o}$, the contribution $\gamma_{\phi}^{\rm vV}$ due to
pair-flip processes can be expressed in terms of the second moment
of the homogenous absorption line, $\langle \delta \omega_{\phi}^2
\rangle \approx T_2^{-2}$, by incorporating the $\tilde{g}$-factors
in the van Vleck analysis \cite{vanvleck48PR}:
\begin{subequations}
\begin{eqnarray}
(\gamma_{\phi}^{\rm vV})^2  \; \approx \; \left[1 - \tanh^2 \left(
\frac{\Delta_{\rm o}}{k_{\rm B} T} \right) \right] \sum_{i \ne j}
\left( \frac{{\cal A}_{yy}^{ij}}{\Delta_{\rm o}} \right)^2,
\label{T2} \\
{\cal A}_{yy}^{ij} = \frac{U_{\mathrm{d}}}{(2 g_e S)^2}
[(2\tilde{g}_{y}^2 + \tilde{g}_{z}^2) {\cal R}^{ij}_{y y} -
(\tilde{g}_{x}^2 - \tilde{g}_{z}^2) {\cal R}^{ij}_{x x}],
\label{A_yy}
\end{eqnarray}
\end{subequations}
with ${\cal R}^{ij}_{\mu \nu} = {\cal V}_{\rm c} (|{\bf r}^{ij}|^2
\delta_{\mu \nu} - 3 r^{ij}_{\mu} r^{ij}_{\nu})/|{\bf r}^{ij}|^5$.
The next term, neglected in Eq. (\ref{T2}), is $\sim O(U_{\rm d} /
\Delta_{\rm o})$. This approach fails when $k_{\rm B}T \ll
\Delta_{\rm o}$, since the line assumes a Lorentzian shape, where
$T_2$ is not related to $\langle \delta \omega_{\phi}^2 \rangle$.

\begin{figure}[b]
\includegraphics[width=8.5cm]{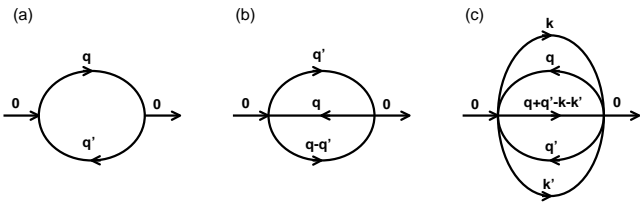}
\caption{Feynman self-energy graphs for the ${\bf q = 0}$ uniform
precession mode, interacting with magnon excitations. We depict (a)
a 3-magnon process, which is ruled out by energy conservation, (b) a
4-magnon process, and (c) a 6-magnon process.}
\label{l-diagrams}
\end{figure}

Consider now magnon-mediated dipolar decoherence. In the experimental set-up we imagine here, a resonant tipping
pulse applied to the spin ensemble causes a subsequent uniform spin precession (ie., coherent tunneling of the
spins in the crystal between states $\vert {\cal Z}_+ \rangle$ and $\vert {\cal Z}_- \rangle$). This is
equivalent, in our effective spin language, to a magnon with wave vector $\mathbf{q}=0$, with gapped energy
$\hbar \omega_{\rm o} \sim 2 \Delta_{\rm o}$ when $\Delta_{\rm o} \gg U_{\rm d}$. However the dipolar interaction
couples this magnon to other magnons \cite{pincus61PR}. The magnon spectrum $\omega_{\mathbf{q}}$ is calculated
using standard Holstein-Primakoff transformations \cite{holstein40PR} applied to Eq. (\ref{H_eff}). The
lowest-order processes conserving both energy and momentum here are 4-magnon processes [cf. Fig.
\ref{l-diagrams}(b)]; for $k_{\rm B} T \ll U_{\rm d}$ these are the only ones that contribute significantly (6-th
and higher-order processes are $\sim O(k_{\rm B} T/U_{\rm d})^2$ relative to these). Selecting the $T_2$ terms
where the total spin polarization is unchanged, we derive a magnon contribution $\gamma_{\phi}^{\rm m}$ to
$\gamma_{\phi}$ given (again, for $\Delta_{\rm o} \gg U_{\rm d}$) by:
\begin{equation}
\gamma_{\phi}^{\rm m} = \frac{2 \pi}{\hbar \Delta_{\rm o}} \sum_{{\bf q q'}}
|\Gamma^{(4)}_{{\bf q} {\bf q'}}|^2 {\cal F}[\overline{n}_{{\bf q}}
]\; \delta(\omega_{\rm o} + \omega_{\bf q} - \omega_{\bf q'} -
\omega_{{\bf q} - {\bf q'}}) \label{2i2o}
\end{equation}
Here the relevant 4-magnon matrix element is $\Gamma^{(4)}_{{\bf q}
{\bf q'}} = (1/4N) [{\cal K}^{(4)}({\bf q}, {\bf q'}) + {\cal
K}^{(4)}({\bf 0}, {\bf q'})]$, where ${\cal K}^{(4)}({\bf q}, {\bf
q'}) = 2 K_{yy}({\bf q}-{\bf q'}) - K_{zz}({\bf q}) - K_{xx}({\bf
q})$ and
\begin{equation}
K_{\mu \nu}({\bf q}) = U_{\rm d} \frac{\tilde{g}_{\mu}
\tilde{g}_{\nu}}{g^2_e S^2} \sum_{l \in V} {\cal R}_{\mu \nu}^{{\it
0} l} e^{ i {\bf q \cdot r}^l}.
 \label{K_k}
\end{equation}
${\cal F}[\overline{n}_{{\bf q}}]$ is the usual Bose statistical weighting of the magnon thermal occupation
numbers $\overline{n}_{{\bf q}}$, and $\hbar \omega_{\bf q} = (A^2_{\bf q} - 4 |B_{\bf q}|^2)^{1/2}$ with $A_{\bf
q} = 2 \Delta_{\rm o} - \frac{1}{4} [2K_{yy}(0) - K_{zz}({\bf q}) - K_{xx}({\bf q})]$ and $B_{\bf q} =
\frac{1}{8}[K_{zz}({\bf q}) - K_{xx}({\bf q}) - 2 i K_{xz}({\bf q})]$ for any $\bf q$. The magnon analysis
requires spin polarization close to unity, ie., $k_{\rm B}T < 2 \Delta_{\rm o}$, and therefore complements the
`van Vleck' approach from the low-$T$ side. It's interesting to notice that the two methods yield different
$T$-dependencies, $\sim \exp(-\Delta_{\rm o}/k_{\rm B}T)$ for $\gamma_{\phi}^{\rm vV}$ and $\sim
\exp(-2\Delta_{\rm o}/k_{\rm B}T)$ for $\gamma_{\phi}^{\rm m}$. The crossover occurs around $k_B T \sim
\Delta_{\rm o}$, where the lineshape turns from Gaussian ($k_{\rm B}T > \Delta_{\rm o}$) to Lorentzian ($k_{\rm
B}T < \Delta_{\rm o}$). Full quantitative details will appear in a long paper \cite{dipD-L}.

\begin{figure}[b]
\includegraphics[width=8.7cm]{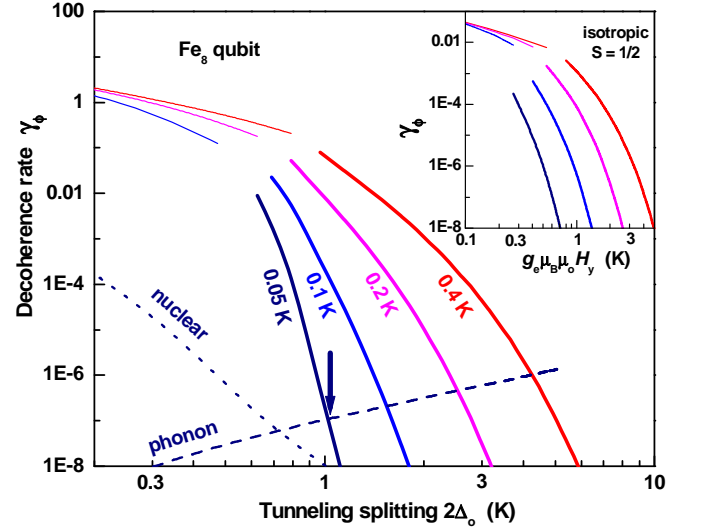}
\caption{(Color online) Dimensionless decoherence rates $\gamma_{\phi} = \hbar / T_2 \Delta_{\rm o}$ as a
function of tunneling gap $2\Delta_{\rm o}$ in Fe$_8$, at the indicated $T$. Thin lines: $\gamma_{\phi}^{\rm vV}$
arising from pair-flip processes, Eq. (\ref{T2}). We omit $\gamma_{\phi}^{\rm vV}$ at $T=0.05$ K $\ll U_{\rm
d}/k_{\rm B}$. Thick lines: $\gamma_{\phi}^{\rm m}$ from magnon scattering, Eq. (\ref{2i2o}). The gap between the
$\gamma_{\phi}^{\rm vV}$ and $\gamma_{\phi}^{\rm m}$ lines is the crossover region between the validity of the
two methods. The dashed and dotted lines show respectively the phonon [$\gamma_{\phi}^{\mathrm{ph}}$, Eq.
(\ref{F_AS})] and nuclear ($\gamma_{\phi}^{\rm NS}$) decoherence rates at $T = 0.05$ K. The arrow indicates the
optimal operation point of the Fe$_8$ spin qubit at $T = 0.05$ K. Inset: $\gamma_{\phi}^{\rm vV}$ and
$\gamma_{\phi}^{\rm m}$ for an isotropic spin-1/2 on the same Fe$_8$ lattice, as a function of the Zeeman gap
$g_e \mu_{\rm B} \mu_{\rm o} H_y$. }
 \label{l-rates}
\end{figure}

To these 2-qubit decoherence processes one must also add 1-qubit
decoherence processes to find the total $\gamma_{\phi}$ that would
be measured in an experiment. The contributions from interaction
with phonons and nuclear spins have been calculated for Fe$_8$
elsewhere \cite{stamp04PRB}. In large transverse fields nuclear
spins give a rate $\gamma_{\phi}^{\mathrm{NS}} = E^2_{\rm o} /
2\Delta_{\rm o}^2$, where $E_{\rm o}$ is the half-width of the
Gaussian multiplet of nuclear-spin states coupled to the qubit. The
phonon decoherence rate is given for Fe$_8$ by \cite{Phon_expl}:
\begin{equation}
\gamma_{\phi}^{\mathrm{ph}} = \frac{ {\cal M}^2_{{\cal A} {\cal S}}
\Delta_{\rm o}^2} {\pi \rho c^5_s \hbar^3} \coth \left( \frac{\Delta_{\rm o}}
{k_{\rm B} T} \right), \label{F_AS}
\end{equation}
where ${\cal M}^2_{{\cal A} {\cal S}}(H_y) \approx \frac{4}{3} D^2
|\langle {\cal A} |S_y S_z + S_z S_y | {\cal S} \rangle|^2$, with
density $\rho = 1920$ kg/m$^3$ and sound velocity $c_s = 1386$ m/s.

Fig. \ref{l-rates} summarizes the results of all these calculations. Except at very low $T$ and large
$\Delta_{\rm o}$, dipolar decoherence completely dominates over nuclear and phonon decoherence. The optimal
operating point at $T = 0.05$ K, having minimum total decoherence, is found where $2 \Delta_{\rm o} \approx 1$ K
$\approx 20$ GHz, at $\mu_{\rm o} H_y \approx 2.6$ T; here the decoherence quality factor is $Q_{\phi} \sim
10^7$, with a coherence time $T_2 \sim 1$ ms. Since $\gamma_{\phi}^{\mathrm{ph}} \propto \coth(\Delta_{\rm
o}/k_{\rm B}T)$ is essentially $T$-independent in this regime, while dipolar contributions to $\gamma_{\phi}$
still vary strongly, a further decrease in temperature would not substantially decrease $\gamma_{\phi}$, but
would allow operation at lower frequencies. It is instructive to compare the results for the Fe$_{8}$ qubit, with
the case of a fictitious isotropic spin-1/2 on the same lattice (inset Fig. \ref{l-rates}), obtained by setting
$\tilde{g}_{x} = \tilde{g}_{y} = \tilde{g}_{z} = g_e$ and replacing $2 \Delta_{\rm o}$ by $g_e \mu_{\rm B}
\mu_{\rm o} H_y$ in all our formulas. We then deal with simple Larmor precession instead of spin tunneling. There
is a clear reduction in $\gamma_{\phi}$, but no trivial proportionality factor, because of the strong variation
of $\tilde{g}_{\mu}(H_y)$ in the Fe$_8$ qubit [Fig. \ref{l-geff}(b)].

We now consider the more general implications of these results. Note
first that even for the set-up considered here, correlated errors
cause large decoherence - we suspect they will even more strongly
affect higher-order entanglement between the qubits. This is in line
with the results in the quantum information literature
\cite{alicki98}, but the concrete calculation here reveals a
surprising feature, viz., that at very low $T$, the contribution of
correlated errors can be made much {\it smaller} then the
single-qubit errors coming from hyperfine and spin-phonon couplings.

Dipolar interactions are dangerous for spin qubit design, but they are hard to screen. One way to reduce their
effects (apart from going to very low $T$ \cite{dipI}) would be to go to lower dimensional spin networks; recent
progress in attaching and assembling nanomagnets on surfaces \cite{cavallini03NL}, or even in chain structures
\cite{clerac02JACS} might then yield viable architectures. Designs in which dipolar inter-qubit couplings can be
made small - eg., low-spin systems like V$_{15}$ or Cr$_7$Ni \cite{chiorescu00PRL} - and where the inter-qubit
couplings responsible for information manipulation can be switched on and off, are clearly favored.

We believe that we have captured the intrinsic decoherence processes
in networks of coupled spin qubits, extending to the case where the
qubit is the low-energy truncation of a larger system. Spin-echo
experiments on well-characterized systems like Fe$_8$ would give a
stringent test of the theory. Perhaps more important, such
experiments would allow exploration of different spin network
architectures, even before the manipulation of individual spins in
such networks becomes possible.

\begin{acknowledgements}
We thank A. Burin, W. N. Hardy, A. Hines, A. J. Leggett, and G. A. Sawatzky for stimulating discussions, and
NSERC and PITP for support.
\end{acknowledgements}


\begin{thebibliography}{10}
\bibitem{moment} {We drop a term $m^S(\mathbf{H}_{\perp})
 = g_e \mu_B (\langle \mathcal{S}
| S_y | \mathcal{S} \rangle + \langle \mathcal{A} | S_y |
\mathcal{A} \rangle)/2$ from the moment, since it contributes only
to the static demagnetisation field.}
\bibitem{Phon_expl}
{This expression corrects Eq. (9) of \cite{stamp04PRB}, from which a
factor $\sim \Delta^2_{\rm o} / \Omega^2_{\rm o}$ was accidentally omitted.}
\end{thebibliography}
\end{document}